\documentclass[twocolumn,showpacs,aps,prl,superscriptaddress]{revtex4}

\usepackage{xspace}
\usepackage{graphicx}
\usepackage{dcolumn}
\usepackage{amsmath}
\usepackage{epsfig}
\usepackage{supertabular}
\usepackage{subfigure}

\renewcommand{\thesubfigure}{\thefigure.\arabic{subfigure}}
\makeatletter
\renewcommand{\@thesubfigure}{\thesubfigure:\space}
\renewcommand{\p@subfigure}{}
\makeatother

\input{babarsym}

\newcommand{\tauxsec}         {\ensuremath{0.919 \pm 0.003 \nb}\xspace}
\newcommand{\lumi}             {\ensuremath{469 \invfb}\xspace}
\newcommand{\lumion}             {\ensuremath{425 \invfb}\xspace}
\newcommand{\lumioff}             {\ensuremath{44 \invfb}\xspace}

\newcommand{\ntaupairerr}         {\ensuremath{ (4.31 \pm 0.03) \times 10^8}\xspace}

 \newcommand{\uleks}   {\ensuremath{ 4.8 \times 10^{-8}}\xspace}
 \newcommand{\ulmuks}   {\ensuremath{ 7.6 \times 10^{-8}}\xspace}

 \newcommand{\effeks}{\ensuremath{9.1\%}\xspace}
 \newcommand{\effmuks}{\ensuremath{6.1\%}\xspace}

 \newcommand{\bkgsysteks}   {\ensuremath{0.59 \pm (0.19 \oplus 0.17)}\xspace}
 \newcommand{\bkgsystmuks}   {\ensuremath{0.30 \pm (0.17 \oplus 0.05)}\xspace}

\newcommand{\uleksfull}   {\ensuremath{{\cal B}(\taueks) < \uleks }\xspace}
\newcommand{\ulmuksfull}   {\ensuremath{{\cal B}(\taumuks) < \ulmuks }\xspace}

  \newcommand{\effclseks}{\ensuremath{9.4\%}\xspace}
  \newcommand{\effclsmuks}{\ensuremath{7.0\%}\xspace}

 \newcommand{\bkgclseks}   {\ensuremath{1.0 \pm 0.4}\xspace}

 \newcommand{\bkgclsmuks}   {\ensuremath{5.3 \pm 2.2}\xspace}

 \newcommand{\obsclseks}   {\ensuremath{1}\xspace}
 \newcommand{\obsclsmuks}   {\ensuremath{2}\xspace}

 \newcommand{\ulclseks}   {\ensuremath{ 3.3 \times 10^{-8}}\xspace}
 \newcommand{\ulclsmuks}   {\ensuremath{ 4.0 \times 10^{-8}}\xspace}

 \newcommand{\ulclsekssens}   {\ensuremath{ 3.0 \times 10^{-8}}\xspace}
 \newcommand{\ulclsmukssens}   {\ensuremath{ 4.8 \times 10^{-8}}\xspace}

\newcommand{\ulclseksfull}   {\ensuremath{{\cal B}(\taueks) < \ulclseks }\xspace}
\newcommand{\ulclsmuksfull}   {\ensuremath{{\cal B}(\taumuks) < \ulclsmuks }\xspace}

 \newcommand{\pidelmom}   {0.6\xspace}
 \newcommand{\pidellooseeff}   {98\%\xspace}
 \newcommand{\pidelloosepimis}   {10\%\xspace}
 \newcommand{\pideltighteff}   {93\%\xspace}
 \newcommand{\pideltightpimis}   {0.1\%\xspace}

 \newcommand{\pidmumom}  {1.4\xspace}
 \newcommand{\pidmulooseeff}   {92\%\xspace}
 \newcommand{\pidmuloosepimis}   {6\%\xspace}
 \newcommand{\pidmutighteff}   {80\%\xspace}
 \newcommand{\pidmutightpimis}   {2\%\xspace}

 \newcommand{\systlumi}   {0.7\%\xspace}
 \newcommand{\systbec}   {less than 0.2\%\xspace}
 \newcommand{\systtrk}   {1.7\% (1.6\%)\xspace}
 \newcommand{\systpid}   {0.4\% (5.1\%)\xspace}
 \newcommand{\systks}   {1.0\%\xspace}

 \newcommand{\systel}   {2.1\%\xspace}
 \newcommand{\systmu}   {5.5\%\xspace}

\newcommand{\taulks}{\ensuremath{\tau^- \to l^- \KS}\xspace}
\newcommand{\taueks}{\ensuremath{\tau^- \to e^- \KS}\xspace}
\newcommand{\taumuks}{\ensuremath{\tau^- \to \mu^- \KS}\xspace}

\newcommand{\deltam}{\ensuremath{\Delta{M_{\tau}}}\xspace}
\newcommand{\deltae}{\ensuremath{\Delta{E_{\tau}}}\xspace}

 \newcommand{\ccbarbkgpi}{\ensuremath{\Dm \to \KS \pim}\xspace}
 \newcommand{\ccbarbkglep}{\ensuremath{\Dm \to \KS \ellm \nu}\xspace}

\def\cls{\ensuremath{CL_{\scriptscriptstyle S}}\xspace}
\def\clsb{\ensuremath{CL_{\scriptscriptstyle S+B}}\xspace}
\def\clb{\ensuremath{CL_{\scriptscriptstyle B}}\xspace}
\def\msqtag{\ensuremath{m^2_{\rm \scriptscriptstyle TAG}}\xspace}
\def\fmomnu{\ensuremath{\hat{p}_{\nu}}\xspace}
\def\fmomtautag{\ensuremath{\hat{p}_{\rm \scriptscriptstyle TAG}}\xspace}
\def\chisqfull{\ensuremath{\chi^2_{\rm \scriptscriptstyle FULL}}\xspace}
\def\kk       {\mbox{\tt KK2f}\xspace}
\def\tauola     {\mbox{\tt Tauola}\xspace}
\def\photos     {\mbox{\tt Photos}\xspace}
\def\geant4     {\mbox{\tt GEANT4}\xspace}
\def\jetset     {\mbox{\tt JETSET}\xspace}
\def\evtgen     {\mbox{\tt EvtGen}\xspace}
\newcommand{\cerenkov}{Cherenkov\xspace}

\newcommand{\gevccgevcc}{\ensuremath{{\mathrm{\,Ge\kern -0.1em V^2\!/}c^4}}\xspace}

\newcommand{\evcc}{\ensuremath{{\mathrm{\,e\kern -0.1em V\!/}c^2}}\xspace}

\newcommand{\uds}      {\ensuremath{uds}\xspace}

\newcommand{\BABARPubYear}     {08}
\newcommand{\BABARPubNumber}  {044}
\newcommand{\SLACPubNumber} {13487}
\newcommand{\LANLNumber}  {xx} %{0707.2981 [hep-ex]}

\def\figurebox#1#2#3{%
    \def\arg{#3}%
    \ifx\arg\empty
    {\hfill\vbox{\hsize#2\hrule\hbox to #2{\vrule\hfill\vbox to #1{\hsize#2\vfill}\vrule}\hrule}\hfill}%
    \else
    {\hfill\epsfbox{#3}\hfill}%
    \fi}

\begin{document}

\preprint{\babar-PUB-\BABARPubYear/\BABARPubNumber} 
\preprint{SLAC-PUB-\SLACPubNumber} 
\preprint{\LANLNumber}

\begin{flushleft}
    \babar-PUB-\BABARPubYear/\BABARPubNumber, SLAC-PUB-\SLACPubNumber\\
    %arXiv:\LANLNumber\\
\end{flushleft}

\title{
{\large \bf \boldmath
Search for Lepton Flavour Violating Decays \taulks with the \babar Experiment
}}

%% author list as of 04-Aug-2008 (520 authors)
%
\author{B.~Aubert}
\author{M.~Bona}
\author{Y.~Karyotakis}
\author{J.~P.~Lees}
\author{V.~Poireau}
\author{E.~Prencipe}
\author{X.~Prudent}
\author{V.~Tisserand}
\affiliation{Laboratoire de Physique des Particules, IN2P3/CNRS et Universit\'e de Savoie, F-74941 Annecy-Le-Vieux, France }
\author{J.~Garra~Tico}
\author{E.~Grauges}
\affiliation{Universitat de Barcelona, Facultat de Fisica, Departament ECM, E-08028 Barcelona, Spain }
% ...
\collaboration{The \babar\ Collaboration}
\noaffiliation

\begin{abstract}
A search for the lepton flavour violating decays \taulks ($l = e$ or $\mu$) has been performed
using a data sample corresponding to an integrated luminosity of \lumi, 
collected with the \babar detector at the SLAC PEP-II \epem asymmetric energy collider.
No statistically significant signal has been observed in either channel and
the estimated upper limits on branching fractions are \ulclseksfull and \ulclsmuksfull 
at 90\% confidence level.
\end{abstract}


\maketitle

In the Standard Model (SM), lepton-flavor-violating (LFV) decays of charged
leptons are forbidden or highly suppressed even if neutrino mixing is taken
into account~\cite{Marciano:1977wx,Lee:1977tib,Cheng:1977nv}.
Any occurrences of LFV decays with measurable branching fractions (BFs) 
would be a clear sign of new physics. 
No signal has been found in extensive searches for LFV in \mmu and \mtau
decays (e.g. $\mu \to e \gamma$~\cite{Brooks:1999pu}, $\tau \to \mu 
\gamma$~\cite{Aubert:2005ye,Aubert:2005wa,Hayasaka:2007vc}). 
However, within the bounds set by searches, some physics models that 
extend the SM include new sizable LFV processes.
For a review, see Ref.~\cite{Raidal:2008jk}.
In this paper a search for \taulks decays is presented~\cite{ftnt1}. 

The \taulks BF has been estimated in SM
extensions with heavy singlet Dirac neutrinos~\cite{Ilakovac:1999md}
and in $R$-parity violating supersymmetric models~\cite{Saha:2002kt}. In
the first case, heavy neutrinos with large mass and large mixing with
SM leptons are introduced.  Because of the large number of independent
angles and phases in the enlarged mixing matrix, the LFV amplitude cannot be
precisely evaluated.
In the large-mass limit of heavy neutrinos and keeping only the
leading terms, theoretical upper bound estimations are of the order $10^{-16}$
and are thus out of experimental reach.
In the second case, couplings of SM leptons to new particles are
described using an $R$-parity violating superpotential.
With many new complex couplings, the phenomenology is immensely richer, but at the
same time less predictive. While $R$-parity conserving couplings can affect
low-energy processes only through loops, $R$-parity violating contributions can
appear as tree-level slepton or squark mediated processes, competing
with SM contributions. 
So, while LFV decays are highly suppressed in the SM, they
can be significantly enhanced in $R$-parity violating supersymmetry.
The previous best experimental upper limits (ULs) for \taulks decay branching ratios were
measured by the Belle Collaboration using a 281
\invfb data sample: ${\cal B}(\taueks)<5.6 \times 10^{-8}$ and ${\cal
B}(\taumuks)<4.9 \times 10^{-8}$ at 90\% confidence level~\cite{Miyazaki:2006sx}.

\par 
The measurement described in this paper is performed using data collected by the
\babar~detector at the PEP-II asymmetric energy storage ring. Charged particles are
detected and their momenta measured by a combination of a silicon
vertex tracker (SVT), consisting of 5 layers of double-sided
detectors, and a 40-layer central drift chamber (DCH), both operating
in a 1.5 T axial magnetic field. Charged particle identification (PID) is
provided by the energy loss in the tracking devices and by the
measured \cerenkov angle from an internally reflecting ring-imaging
\cerenkov detector (DIRC) covering the central region. Photons are
measured, and electrons detected, by a CsI(Tl) electromagnetic calorimeter (EMC).
The EMC is surrounded by an instrumented flux return (IFR).  
The detector is described in detail elsewhere~\cite{Aubert:2001tu,Menges:2006xk}.

The analyzed data sample corresponds to an integrated luminosity of \lumi
collected from \epem collisions, \lumion at
the $\Upsilon(4S)$ resonance and \lumioff at center-of-mass (CM)
energy 10.54\gev.
The total number of produced \mtau pairs $N_{\tau \tau}$ is \ntaupairerr, 
calculated using the average \mtau cross section of \tauxsec estimated with 
\kk~\cite{Banerjee:2007is}.
The Monte Carlo (MC) simulated samples of \mtau leptons
are produced using the \kk generator~\cite{Jadach:1999vf,Ward:2002qq} 
and \tauola decay library~\cite{Jadach:1993hs,Barberio:1993qi}.
Decays of $B$ mesons are simulated with the \evtgen
generator~\cite{Lange:2001uf}, while $\epem \to q\bar{q}$ events, where $q=u,d,s$ quarks 
(referred to as \uds events) or $q=c$ quark, are simulated with the \jetset
generator~\cite{Sjostrand:2006za}. The \babar detector is
modeled in detail using the \geant4 simulation package~\cite{Agostinelli:2002hh}.
Radiative corrections for signal and background processes are simulated using 
\photos~\cite{Golonka:2003xt}. In the following, the simulated signal and 
background samples will be referred to as signal MC and background MC samples, respectively.

For this analysis, two different stages of selection are used.  In the
first, which we call the loose selection stage, we retain enough data to
estimate background distribution shapes. The second, which we refer to
as the tight selection, uses criteria that have been chosen to optimize
the sensitivity. The sensitivity, or expected UL, is defined as the UL 
value obtained using the background expected from MC:
we choose selection criteria that give the smallest expected UL.
We use loose and tight electron and muon PID selectors for the two stages of selection.
The selectors are based on combinations of measurements from the various subdetectors.
The average efficiency for the loose electron (muon) selector is 
\pidellooseeff (\pidmulooseeff) for a laboratory momentum 
$p_{\rm LAB}>\pidelmom\:(\pidmumom)\gevc$, whereas the $\pi$ misidentification 
rate is less than \pidelloosepimis (\pidmuloosepimis). 
The average identification efficiency for the tight electron (muon) 
selector with a likelihood based algorithm is \pideltighteff (\pidmutighteff) for 
the same momentum range, whereas the $\pi$ misidentification rate is less than 
\pideltightpimis (\pidmutightpimis). 
All selection criteria are applied to both channels and
quantities are defined in the CM system, unless stated otherwise.

Events are first selected using global event properties in order to reject \bbbar, \ccbar
and \uds background events with high multiplicity. 
All tracks (photons) are required to be reconstructed within a 
fiducial region defined by $0.410 < \theta < 2.540 \; (0.410 < \theta < 2.409)$ radians, where 
$\theta$ is the polar angle in the laboratory system
with respect to the $z$ axis direction~\cite{Aubert:2001tu}.
The overall event charge must be zero.
Furthermore, the event must include a \KS candidate 
with an invariant mass within 25\mevcc of the nominal \KS mass~\cite{Yao:2006px},
reconstructed from two oppositely charged tracks, assuming the pion mass for both.
The highest momentum track in the CM frame has to have a momentum between 1.5  
and 4.8\gevc for both modes.
For the electron channel events, the total EMC energy associated 
with tracks in the laboratory frame has to be less than 9\gev.
The thrust~\cite{Brandt:1964sa} is calculated using tracks and calorimeter energy
deposits without an associated charged particle track.
The thrust magnitude has to be between 0.85 (0.88) and 0.98 (0.97) for the 
electron (muon) channel.
For each event, two hemispheres are defined in the 
CM frame using the plane perpendicular to the thrust axis.
The hemisphere that contains the reconstructed \mtau candidate, defined below,
is referred to as the signal side and the other
hemisphere as the tag side. 
Candidate \mtau pair events are required to have three reconstructed 
charged particle tracks on the signal side.
On the tag side, one track only is required for the muon channel, while for 
the electron channel, events with one or three reconstructed tracks are retained.

The signal \mtau candidates are reconstructed by combining 
one \KS candidate with the third track of the signal hemisphere,
to which mass is assigned according to the considered decay mode.
The lepton track is required to be identified as an electron 
or muon by the loose PID selector.
The signal \mtau candidates are then examined in the two 
dimensional distribution of \deltae vs.\@ \deltam, where
\deltam is defined as the difference between the invariant mass of the
reconstructed \mtau and the world average value~\cite{Yao:2006px}, and \deltae is 
defined as the difference between the energy of the reconstructed \mtau 
and the expected \mtau energy, half the CM total energy.
Only \mtau candidates with a \deltam value within $\pm0.35$ \gevcc 
and a \deltae value within $\pm0.4$ \gev are retained. 
The whole decay tree is then fitted requiring that, 
within reconstruction uncertainties, 
the \KS decay products form a vertex,
the \KS mass is constrained to the nominal value, 
and the track and the \KS trajectory form a vertex close to the
beam interaction region.
To improve the energy resolution, a bremsstrahlung recovery procedure is applied for the
\taueks decay mode only: before the fit, the \en track candidate is combined 
with up to three photons with an energy
larger than 30\mev and contained in a cone around the track
direction of $\Delta\theta\times\Delta\phi=0.035\times0.050\rad^2$,
where $\theta$ is the polar angle and $\phi$ the azimuthal angle in 
the laboratory system.
The constrained fit must have a $\chi^2$ probability 
larger than 1\%. If more than one candidate is found 
(which occurs in less than 1\% of the events), only that with
the largest $\chi^2$ probability is retained.

After the above selection is applied, backgrounds remain, mainly from 
Bhabha events for the electron channel and from non-lepton events for 
the muon channel due to the larger pion to muon misidentification. To 
improve the background rejection, further requirements are imposed on the \KS 
candidates. For the muon channel, the \KS laboratory momentum must be 
greater than 1.0\gevc. For the electron channel, in order to remove 
events with a photon conversion faking a \KS, the invariant mass of the 
\KS daughters, calculated using the momentum from the fit and assigning 
them the electron mass, is required to be greater than 0.10\gevcc. The 
\KS flight length significance is computed as the three-dimensional 
distance in the laboratory system between the \mtau vertex and the \KS 
vertex, divided by its error, and we select events with a flight length 
significance greater than 3.0. Finally, the \KS reconstructed mass is 
required to be between 0.482 and 0.514\gevcc. The last two criteria are 
included in the loose selection for the electron channel while, for the 
muon channel, they are applied at a later stage in order to maintain 
sufficient statistics in the loose selection sample. 
The amount of background events due to dimuon and Bhabha processes 
is negligible after the loose selection has been applied and
most of the surviving events come from charm decays, 
such as \ccbarbkgpi and \ccbarbkglep, and from
combinations in the \uds events of a true \KS and a fake lepton.

\begin{figure}[!htbp] 
 \begin{center}
   \includegraphics[width=8cm]{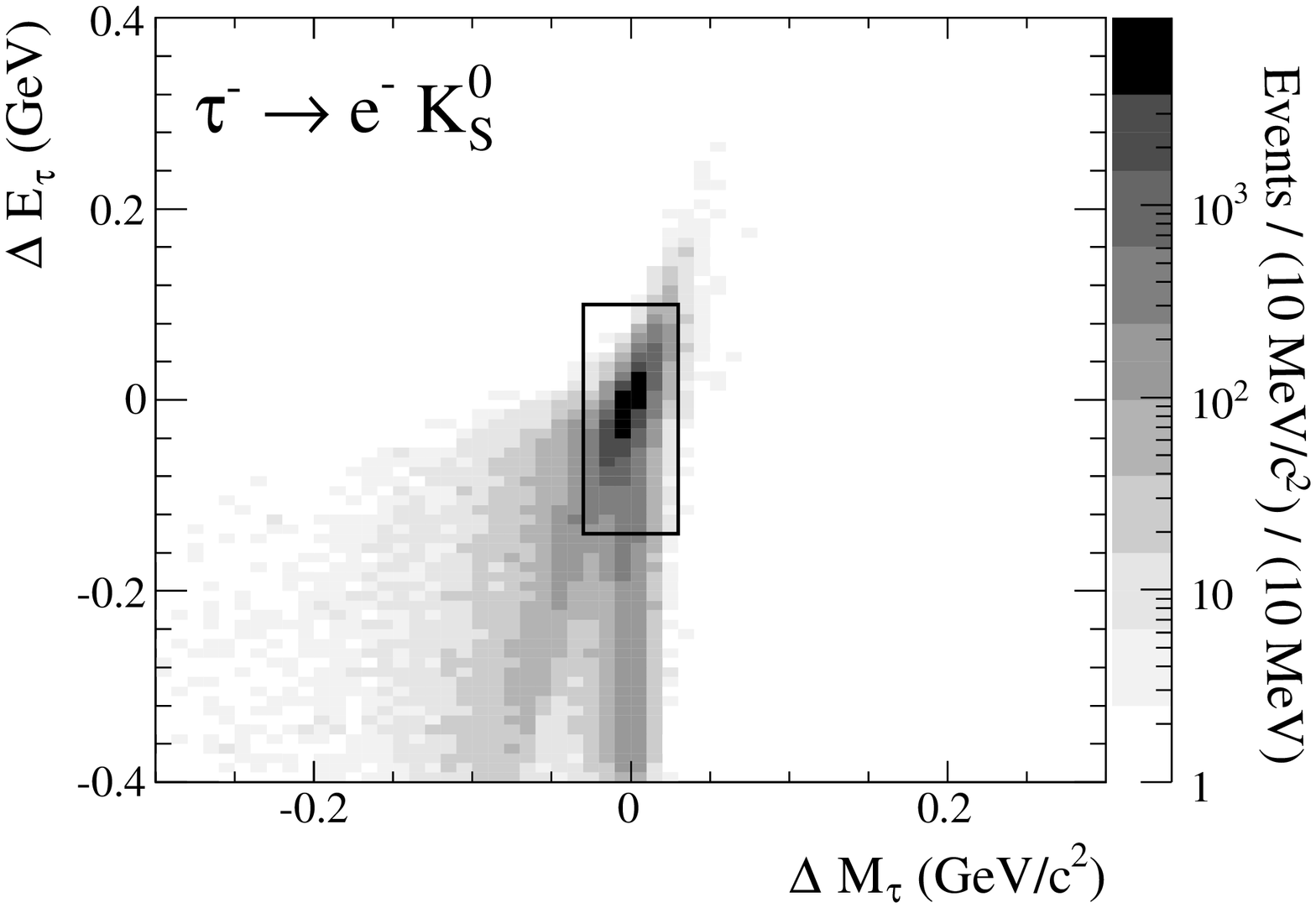}
    \vspace{-.2cm}
   \includegraphics[width=8cm]{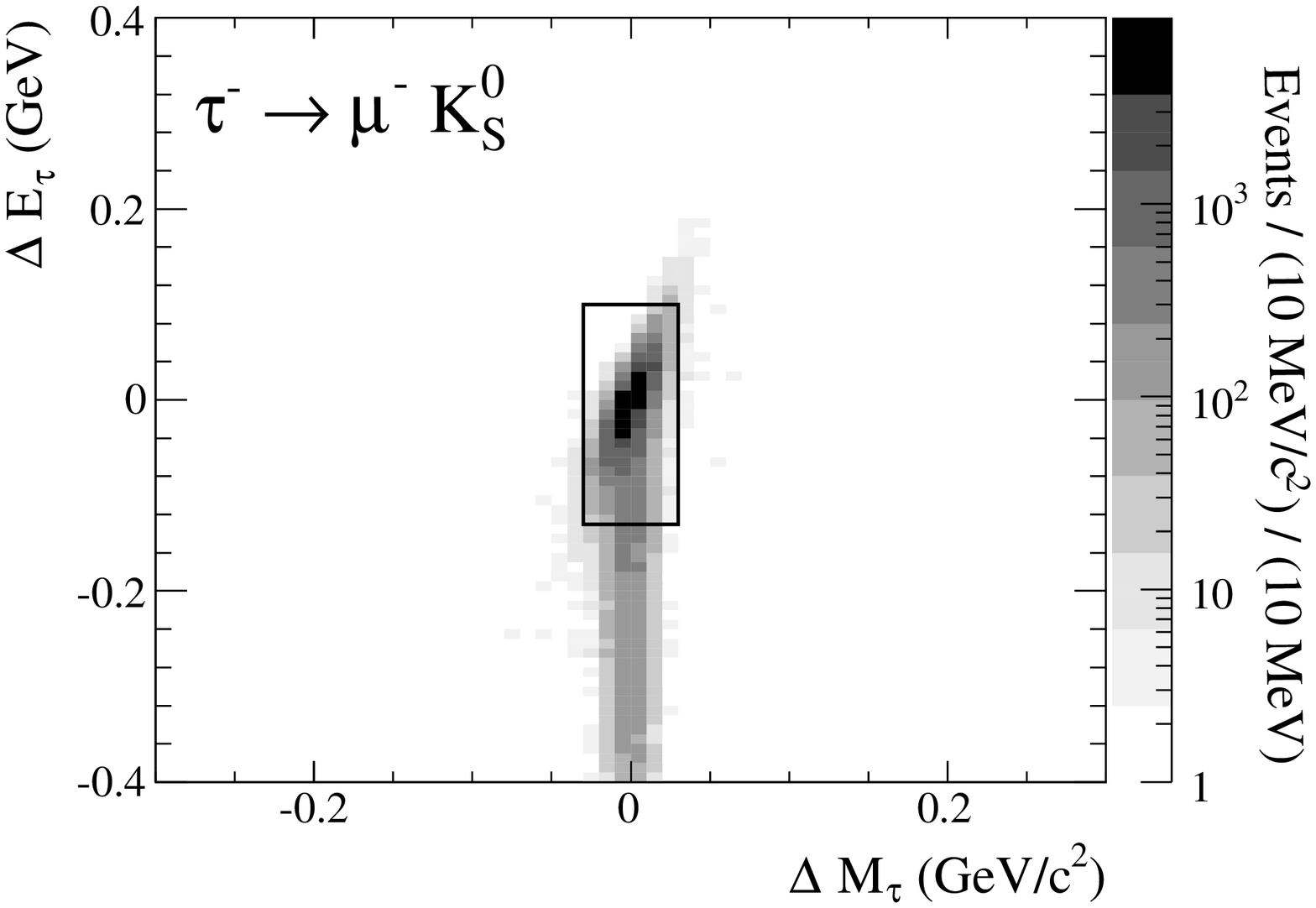}
    \vspace{-.2cm}
    \caption{Candidate distributions for signal MC samples \taueks (top) 
      and \taumuks (bottom) in the (\deltae, \deltam) 
      plane after the loose selection. The rectangle corresponds to the blinded box.
      The $z$-axis scale is logarithmic.
    }
    \label{fig:dedm_loose}
  \end{center}
    \vspace{-.6cm}
\end{figure}

\begin{figure}[!htbp] 
 \begin{center}
   \includegraphics[width=8cm]{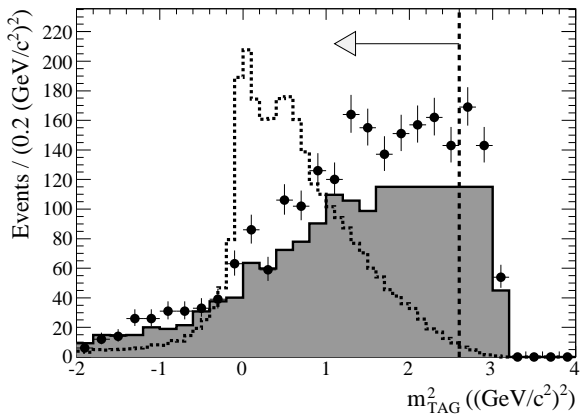}
    \vspace{-.4cm}
    \caption{Distributions of \msqtag after the loose selection 
      for the \taueks channel.
      The data distribution is shown by solid circles with error bars, background MC 
      with a filled histogram
      and signal MC with a dashed line.
      The signal MC distribution is normalized arbitrarily, while the background MC is normalized 
      to the data luminosity. The vertical dashed line and the arrow indicate the applied requirement.}
    \label{fig:msqtag_loosesel}
    \vspace{-.5cm}
  \end{center}
\end{figure}

To avoid bias from adapting selection requirements to the data,
the tight selection has been optimized in a blind way, without looking at the data
in the rectangular region (blinded box) shown in Fig.~\ref{fig:dedm_loose},
corresponding to more than $\pm 5$ times the resolution for signal events
on \deltae and \deltam, respectively. 
As discussed above, selection criteria have been chosen to optimize the sensitivity on the upper limit.
Therefore, for the tight selection the tighter PID selectors 
plus the following requirements are applied.
The event's missing momentum is computed by subtracting from the \epem
momentum all track candidates and all unmatched calorimeter energy deposits.
To reject events with tracks and photons lost out of the acceptance,
the missing momentum is required to have a transverse
component greater than 0.1 (0.2)\gevc for the electron (muon) channel
and the cosine of its polar angle in the laboratory system must be smaller than 0.95.
In a \mtau pair event, when neglecting radiation, the tag-side \mtau has
the same momentum as the signal-side \mtau but the opposite direction.
In addition, assuming that the tag \mtau decays to a one neutrino (hadronic) mode,
the event's missing momentum corresponds to the neutrino momentum. 
These two assumptions determine the tag \mtau 4-momentum \fmomtautag, and
the neutrino 4-momentum \fmomnu, respectively, and we define the
squared invariant mass \msqtag as $(\fmomtautag - \fmomnu)^2$.  
As shown in Fig.~\ref{fig:msqtag_loosesel}, \msqtag
peaks at small values
for signal events
and extends to higher values for background events. 
The tail on the right for the signal sample is due to tag \mtau decays
to (leptonic) modes with two neutrinos, while the tail
on the left for the background sample is due
to events with missing energy from lost photons or tracks.
The variable \msqtag is required to be smaller than 2.6 $(\gevcc)^2$ for both channels.
Shapes for data and MC agree within error but a discrepancy is observed in the normalization.
This does not affect the results because the final number of background events 
is obtained using the data sample.
The \uds background events are further reduced by requiring less than six photons on the tag side.
Signal events have missing momentum due only to
the undetected neutrino(s) from the tagging \mtau decay. 
Therefore, only for the \taumuks channel, 
the cosine of the angle between the missing momentum and the signal \mtau 
candidate is required to be negative, 
to further reject non-leptonic backgrounds and improve the sensitivity.

\begin{figure}[htbp] 
 \begin{center}
   \includegraphics[width=8cm]{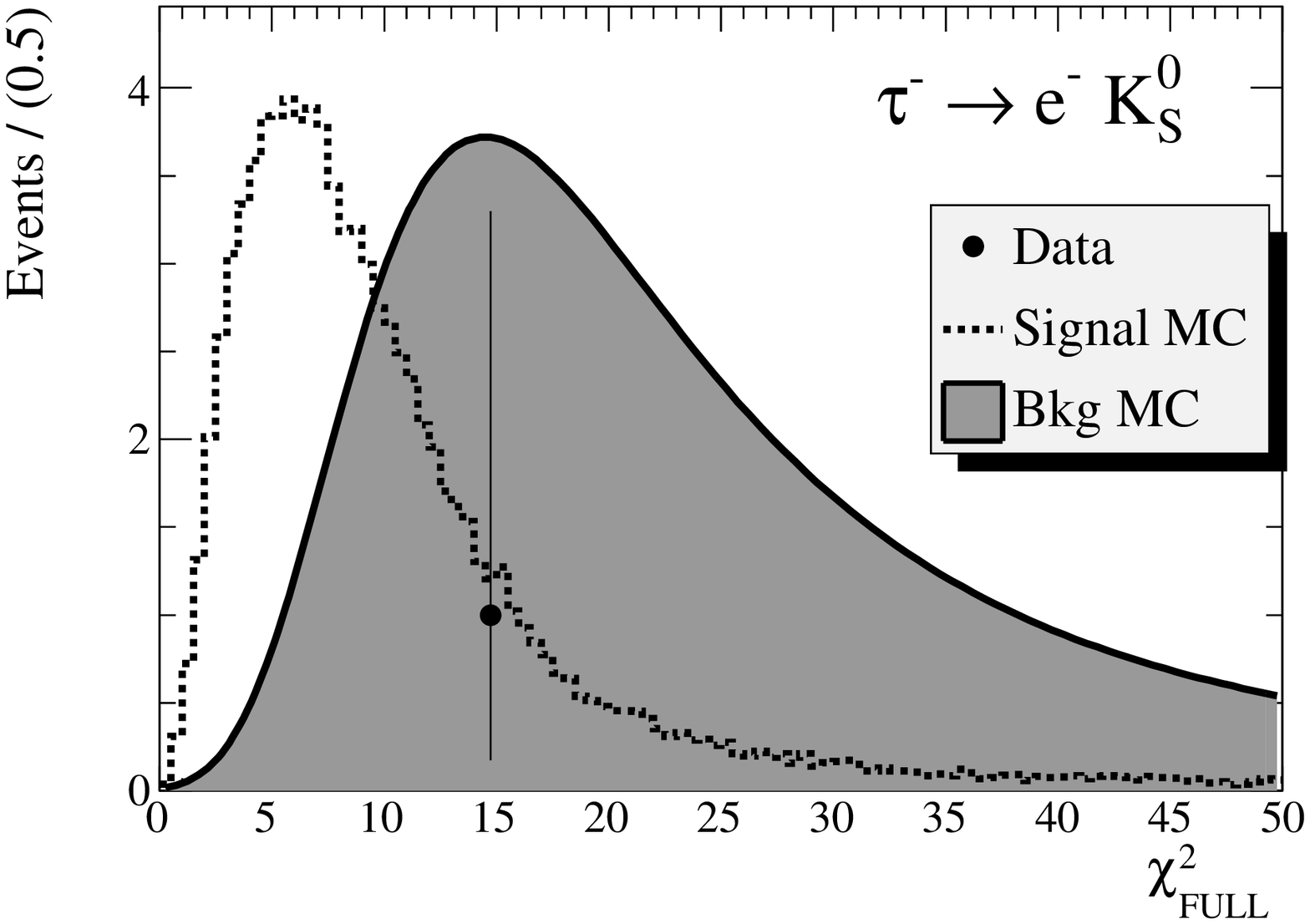}
    \vspace{-.2cm}
   \includegraphics[width=8cm]{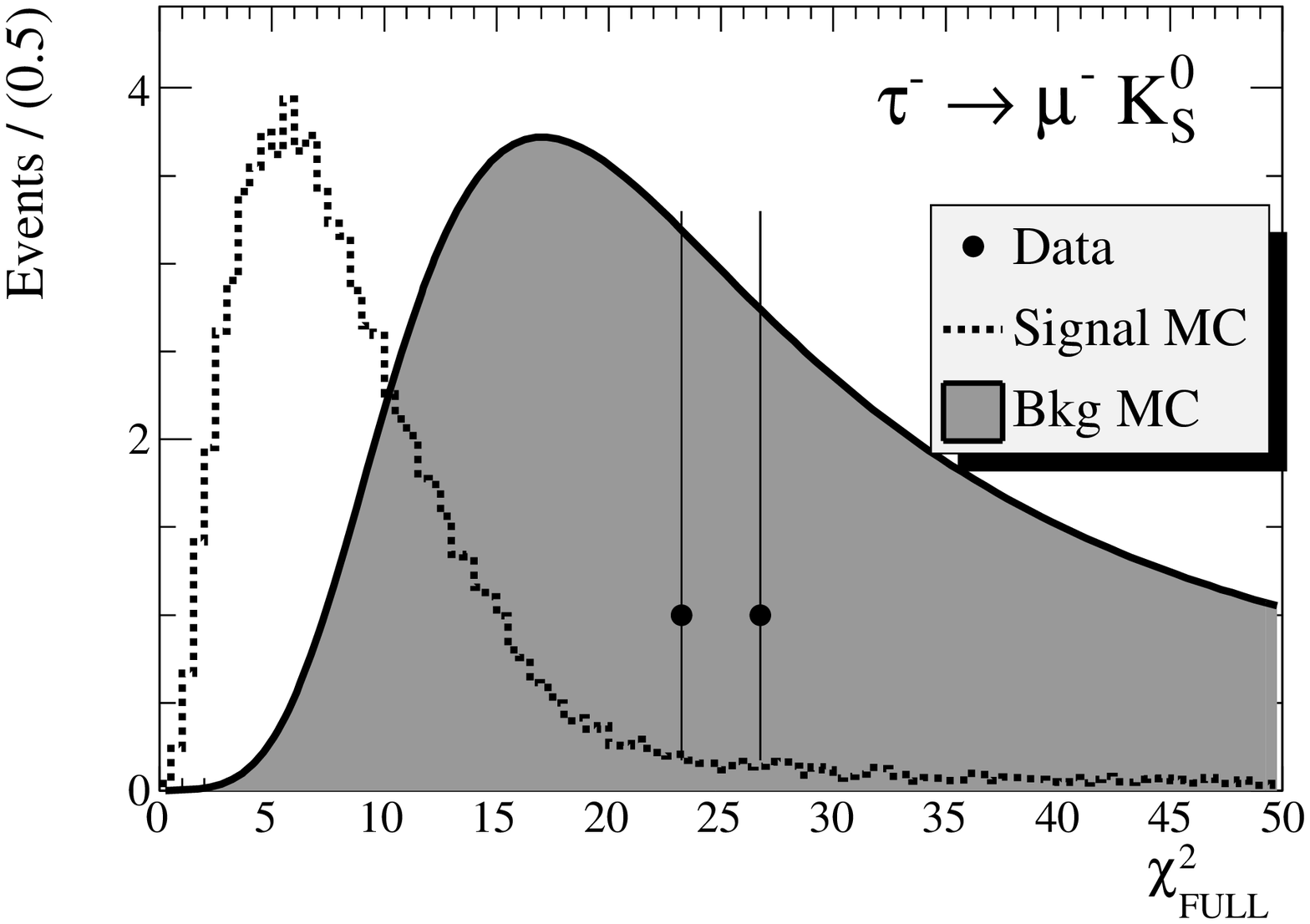}
    \vspace{-.2cm}
    \caption{Distributions of \chisqfull after the tight selection 
      for the \taueks (top) and \taumuks (bottom) channel.
      The data events are shown by solid circles with error bars.
      The signal MC distributions are shown by dashed lines, while
      the background shapes are shown with filled histograms.
      The signal and background MC distributions are normalized arbitrarily.
    }
    \label{fig:chi2full}
  \end{center}
    \vspace{-.6cm}
\end{figure}

For the final step of analysis, we define another discriminating variable,
\chisqfull, as the $\chi^2$ of the geometrical and kinematical fit for 
the whole decay tree, with additional constraints of \deltam and \deltae equal to 0.
Most signal events have \chisqfull values in the range 0-50,
and we consider this range in the following.
In Fig.~\ref{fig:chi2full} we show the distributions of \chisqfull for data and
signal MC inside the blinded box after the tight selection.
An analytic curve describing the background, as detailed in the following, is also
presented. 
The overall efficiency $\varepsilon$ in this range of \chisqfull, 
after the tight selection, and inside the blinded box 
is \effclseks for the \taueks mode and \effclsmuks for the \taumuks mode. 
The total signal efficiency is estimated by dividing the number of 
selected signal MC events by the total number of generated \taulks decays 
and includes the $\KS \to \pipi$ BF.

We estimate the number of background events in the signal region using
the number of MC background events in the range 0-50 of \chisqfull after
the loose selection multiplied by the ratio of numbers of MC background
events after tight and loose selections in the full range of \chisqfull.
We apply a 10\% correction to normalize the MC to the levels 
of background seen in data outside the blinded box after the tight selection.
Total backgrounds of \bkgclseks and \bkgclsmuks events are expected for \taueks
and \taumuks respectively.   
Finally the signal region is unblinded and \obsclseks and \obsclsmuks events 
are found for electron and muon modes, respectively, as already shown
in Fig.~\ref{fig:chi2full}.

\begin{figure}[htbp] 
   \begin{center}
     \includegraphics[width=8.5cm]{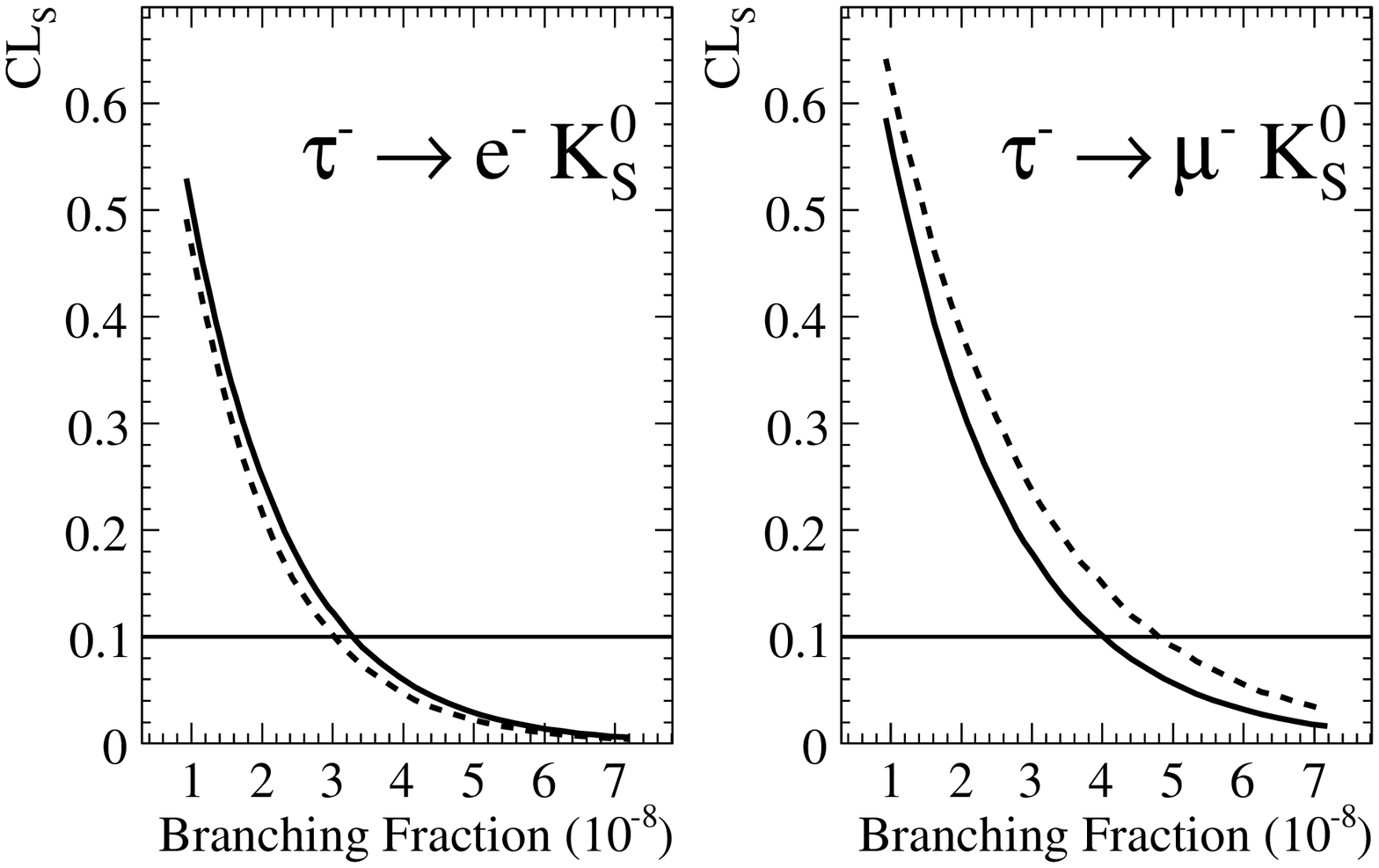}
     \caption{Observed (full line) and expected (dashed line) \cls as a function 
       of the BFs ($10^{-8}$) for the decays \taueks and \taumuks.}
     \label{fig:obsexpcls}
   \end{center}
   \vspace{-.6cm}
\end{figure}

Since no excess above the expected background level is found,
90\% confidence level limits have been determined
according to the modified frequentist analysis (or \cls method)~\cite{Junk:1999kv,James:2000et}. 
This method is more powerful than a simple UL estimation based on numbers of observed and expected events
as it takes into account the different distributions of one or more discriminating variables
between signal and background.
The discriminating variable used in this analysis is \chisqfull.
The signal \chisqfull distribution is simply provided by the MC sample as already shown 
in Fig.~\ref{fig:chi2full}, but this cannot be done for the background as too 
few events survive the tight selection, but also the loose one.
Therefore we obtain smooth background shapes by fitting the product 
of a Landau function and a straight line to the MC background 
distributions after the loose selection.
Any distortions on the shapes that could be introduced by the tight selection
are negligible compared to the uncertainties of the shapes themselves.
The resulting curves are presented in Fig.~\ref{fig:chi2full}.
The adopted test-statistic is the likelihood ratio 
$Q = {\mathcal L}(S+B)/{\mathcal L}(B)$, where ${\mathcal L}(B)$ and 
${\mathcal L}(S+B)$ are, respectively, the likelihood to find the 
observed events in the hypothesis of background only and of background plus 
a given amount of signal. The latter, and consequently $Q$, 
are functions of the hypothesized signal BF.
The confidence level \cls is defined as the ratio $\clsb/\clb$,
where \clsb and \clb are estimated using an ensemble of simulated
datasets, generated from signal plus background or background only.
The generation is iterated with a varying hypothetical value of the
number of signal events, depending on the BF. \clsb and \clb are then
the probabilities that the test-statistic would be less than the
$Q_{exp}$ values observed in data, under the respective hypothesis.
Signal hypotheses corresponding to $\cls < \alpha$ are rejected 
at the $1-\alpha$ confidence level. 
This method 
avoids that a negative fluctuation of the background is translated 
into a large improvement of the exclusion limit and allows to include 
uncertainties directly on signal and background distributions.
The ULs on BFs at 90\% confidence level are calculated as
 \begin{equation}
   \label{eq:ul_bf}
   {\cal B}(\taulks) < \frac{s_{\rm 90}}{2 \varepsilon N_{\tau \tau}}
 \end{equation}
where $s_{\rm 90}$ is the limit for the signal yield at 90\% confidence level, and 
$\varepsilon$ and $N_{\tau \tau}$ are already defined above.
The dominant systematic uncertainties on the signal efficiency for the electron (muon)
channel come from possible data/MC differences in the efficiency of the PID requirements,
\systpid and of the tracking reconstruction, \systtrk.
Other sources of systematic uncertainty for the efficiency are:
data/MC differences in \KS reconstruction efficiency (\systks),
the beam energy scale and the energy spread (\systbec). 
The efficiency errors from MC statistics are negligible compared with the systematics ones.
The uncertainty for the total number of \mtau pairs comes from
the error on the luminosity and on the \mtau cross section values (\systlumi).
We assume these uncertainties are uncorrelated and combine them
in quadrature to give a total signal uncertainty of 
\systel and \systmu respectively for the electron and muon channels. 
For each bin of the signal \chisqfull distribution, we consider the total 
uncertainties on the signal yield, and for the background distributions the 
uncertainties on the expected background levels.
The uncertainties are treated as fully correlated between the bins as they are mainly 
due to normalization uncertainties. 
The analysis results are summarized in Fig.~\ref{fig:obsexpcls} presenting
\cls for the observed events versus the BFs, 
with the horizontal line defining the UL at 90\% confidence level. 

From Fig.~\ref{fig:obsexpcls}, the ULs on the BFs
at 90\% confidence level are determined to be:
\ulclseksfull and \ulclsmuksfull.
The \cls obtained using the number of expected background MC events, instead
of data, are shown in the same figure and the BF values at 90\% confidence level
can be regarded as the sensitivities: 
\ulclsekssens for the electron channel and \ulclsmukssens for the muon one.

\begin{figure}[htbp] 
 \begin{center}
   \includegraphics[width=8cm]{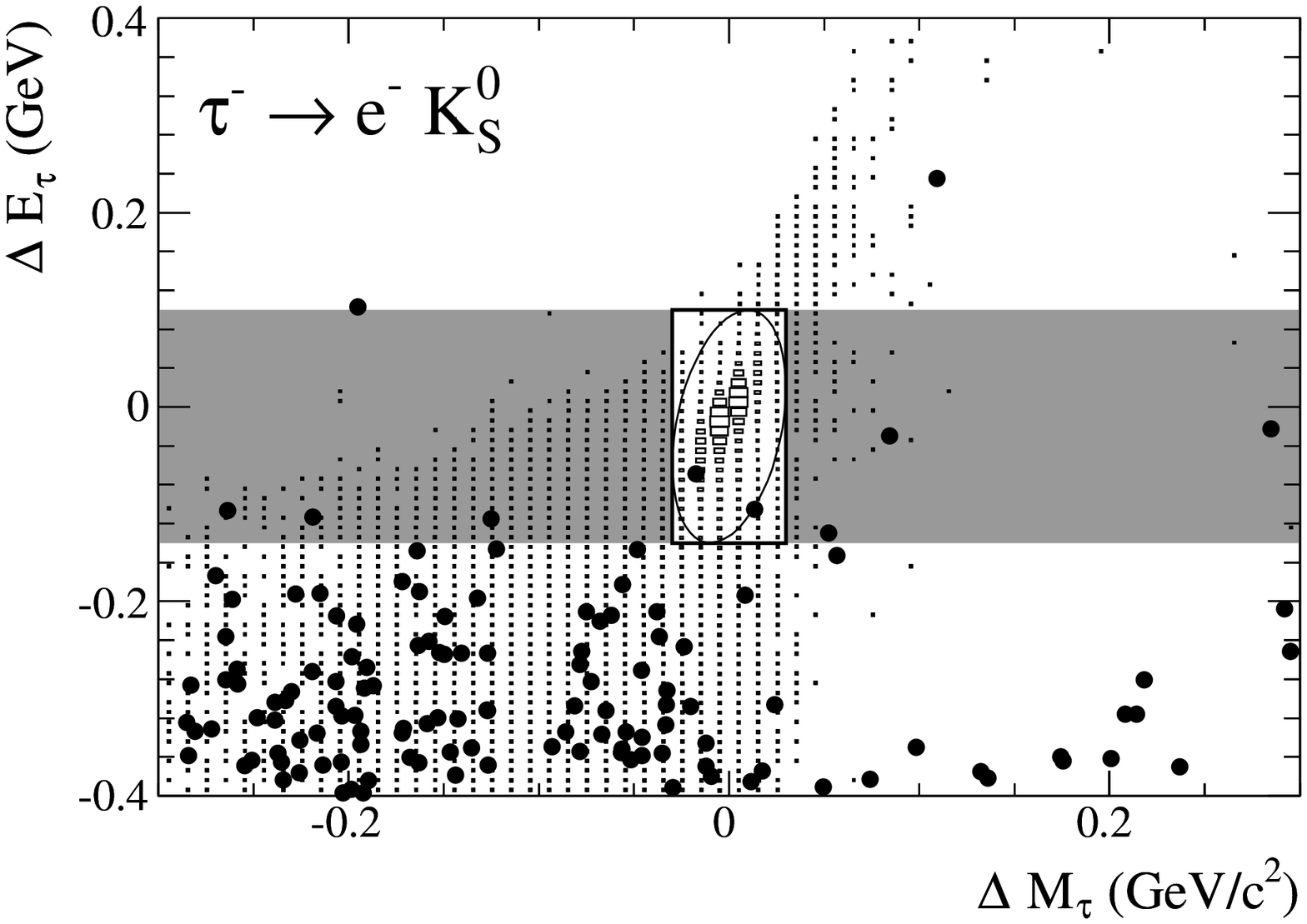}
    \vspace{-.2cm}
   \includegraphics[width=8cm]{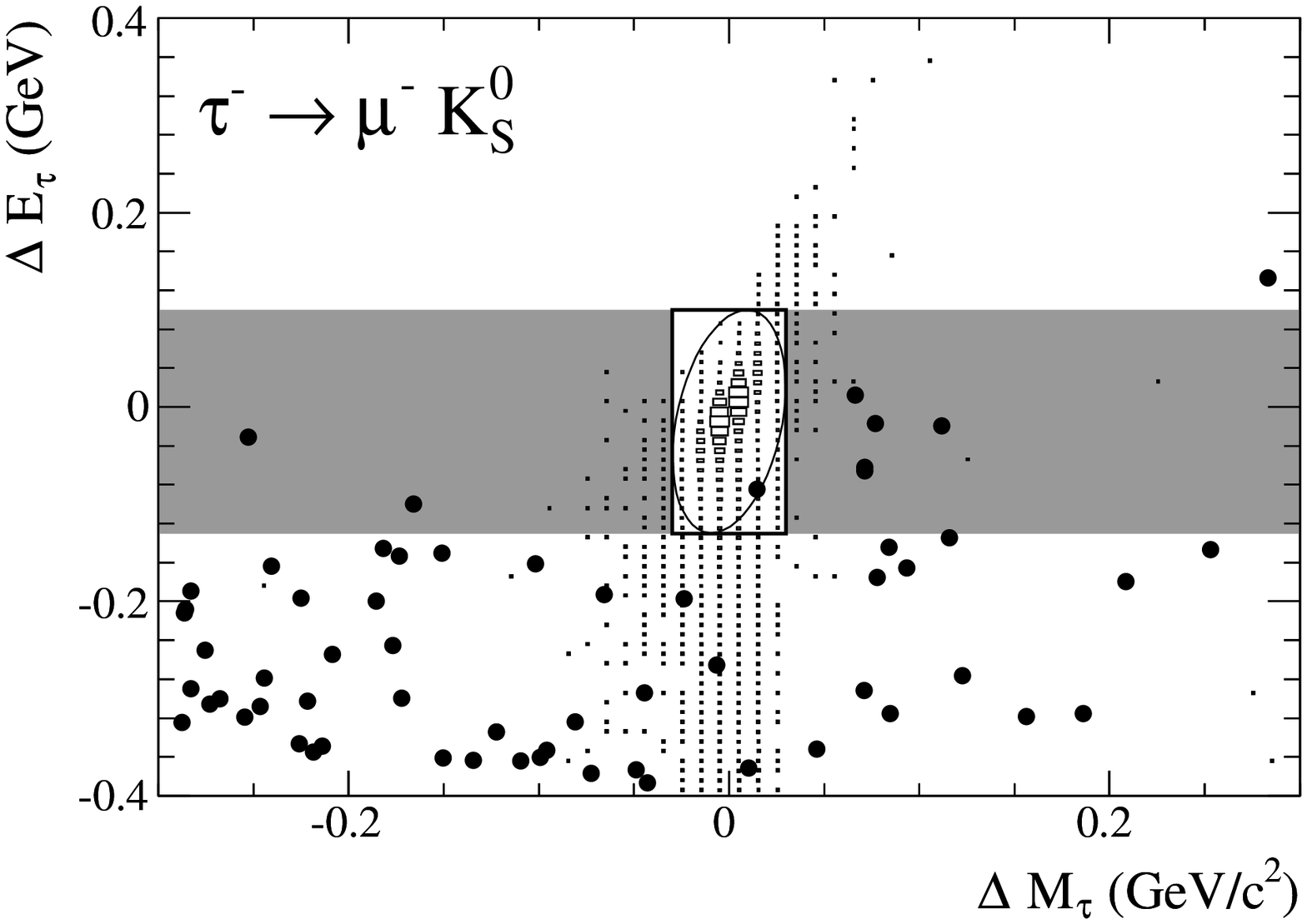}
    \vspace{-.2cm}
    \caption{Candidate distribution in the (\deltae, \deltam) plane after all 
      selections for cross-check method (\taueks on the top, \taumuks on the
      bottom). Data candidates are indicated by solid circles.
      The boxes show the signal MC distribution with arbitrary normalization.
      The blinded box, used for both methods, corresponds to the rectangle.
      The gray bands and the ellipse 
      indicate the sidebands used for extrapolating the background and the signal region
      for the cross-check measurement.
      The $z$-axis scale is linear.
    }
    \label{fig:dedm_final}
  \end{center}
    \vspace{-.6cm}
\end{figure}

ULs are also determined by exploiting another technique that 
gives a similar but worse sensitivity for the UL, so it is used only as cross-check.
For this method, selection criteria on the same quantities
were slightly tightened to reduce the background
as much as possible, and 
signal candidates are counted inside the elliptical
region shown in Fig.~\ref{fig:dedm_final}.
The final signal efficiencies with these selections are 
\effeks for the \taueks mode and \effmuks for the \taumuks mode.
The level of background in the signal ellipse is estimated by
extrapolating the event densities found in two sideband regions of \deltam,
as defined in Fig.~\ref{fig:dedm_final}.
The \deltam background distribution is modeled as a linear function 
plus a Gaussian function to account for the peak related 
to the decay mode \ccbarbkgpi. This is fitted
at the loose selection stage, where there are
sufficient statistics in the sidebands to estimate the shape.
Then the fitted background distribution is normalized according to the
number of data events in the sidebands after the tight selection.
The final estimated number of background events in the signal region is
\bkgsysteks and \bkgsystmuks 
for the electron and muon channels respectively, where the last number is the 
systematic uncertainty accounting for the observed differences between 
estimated and real MC sample events inside the signal region
at the loose selection stage.
When the signal region is unblinded, we find inside the elliptical 
signal region only one event for each channel.
Using the signal efficiencies, the estimated residual backgrounds,
and the number of observed events 
ULs on the BFs at 90\% confidence level for this cross-check
are calculated with the POLE program~\cite{Conrad:2002kn}.
Uncertainties are included assuming that efficiency and
background values have a Gaussian distribution, and that they are not correlated.
The resulting ULs are \uleksfull and \ulmuksfull.

In conclusion, a search for the lepton flavour violating decays \taulks has been
performed using a data sample of \lumi collected with the \babar detector 
at the SLAC PEP-II electron-positron storage rings. 
No statistically significant excess of events is observed in either channel and
the resulting ULs are \ulclseksfull and \ulclsmuksfull at 90\% confidence level.
These results are the most restrictive ULs on the BFs
of these decay modes, and can be used to constrain parts of
the theoretical phase space in several models of physics beyond the
Standard Model.

We are grateful for the 
extraordinary contributions of our \pep2\ colleagues in
achieving the excellent luminosity and machine conditions
that have made this work possible.
The success of this project also relies critically on the 
expertise and dedication of the computing organizations that 
support \babar.
The collaborating institutions wish to thank 
SLAC for its support and the kind hospitality extended to them. 
This work is supported by the
US Department of Energy
and National Science Foundation, the
Natural Sciences and Engineering Research Council (Canada),
the Commissariat \`a l'Energie Atomique and
Institut National de Physique Nucl\'eaire et de Physique des Particules
(France), the
Bundesministerium f\"ur Bildung und Forschung and
Deutsche Forschungsgemeinschaft
(Germany), the
Istituto Nazionale di Fisica Nucleare (Italy),
the Foundation for Fundamental Research on Matter (The Netherlands),
the Research Council of Norway, the
Ministry of Education and Science of the Russian Federation, 
Ministerio de Educaci\'on y Ciencia (Spain), and the
Science and Technology Facilities Council (United Kingdom).
Individuals have received support from 
the Marie-Curie IEF program (European Union) and
the A. P. Sloan Foundation.


\begin{thebibliography}{28}
\expandafter\ifx\csname natexlab\endcsname\relax\def\natexlab#1{#1}\fi
\expandafter\ifx\csname bibnamefont\endcsname\relax
  \def\bibnamefont#1{#1}\fi
\expandafter\ifx\csname bibfnamefont\endcsname\relax
  \def\bibfnamefont#1{#1}\fi
\expandafter\ifx\csname citenamefont\endcsname\relax
  \def\citenamefont#1{#1}\fi
\expandafter\ifx\csname url\endcsname\relax
  \def\url#1{\texttt{#1}}\fi
\expandafter\ifx\csname urlprefix\endcsname\relax\def\urlprefix{URL }\fi
\providecommand{\bibinfo}[2]{#2}
\providecommand{\eprint}[2][]{\url{#2}}

\bibitem[{\citenamefont{Marciano and Sanda}(1977)}]{Marciano:1977wx}
\bibinfo{author}{\bibfnamefont{W.~J.} \bibnamefont{Marciano}} \bibnamefont{and}
  \bibinfo{author}{\bibfnamefont{A.~I.} \bibnamefont{Sanda}},
  \bibinfo{journal}{Phys. Lett. B} \textbf{\bibinfo{volume}{67}},
  \bibinfo{pages}{303} (\bibinfo{year}{1977}).

\bibitem[{\citenamefont{Lee and Shrock}(1977)}]{Lee:1977tib}
\bibinfo{author}{\bibfnamefont{B.~W.} \bibnamefont{Lee}} \bibnamefont{and}
  \bibinfo{author}{\bibfnamefont{R.~E.} \bibnamefont{Shrock}},
  \bibinfo{journal}{Phys. Rev. D} \textbf{\bibinfo{volume}{16}},
  \bibinfo{pages}{1444} (\bibinfo{year}{1977}).

\bibitem[{\citenamefont{Cheng and Li}(1977)}]{Cheng:1977nv}
\bibinfo{author}{\bibfnamefont{T.-P.} \bibnamefont{Cheng}} \bibnamefont{and}
  \bibinfo{author}{\bibfnamefont{L.-F.} \bibnamefont{Li}},
  \bibinfo{journal}{Phys. Rev. D} \textbf{\bibinfo{volume}{16}},
  \bibinfo{pages}{1425} (\bibinfo{year}{1977}).

\bibitem[{\citenamefont{Brooks et~al.}(1999)}]{Brooks:1999pu}
\bibinfo{author}{\bibfnamefont{M.~L.} \bibnamefont{Brooks}}
  \bibnamefont{et~al.} (\bibinfo{collaboration}{MEGA} Collaboration),
  \bibinfo{journal}{Phys. Rev. Lett.} \textbf{\bibinfo{volume}{83}},
  \bibinfo{pages}{1521} (\bibinfo{year}{1999}).

\bibitem[{\citenamefont{Aubert et~al.}(2005)}]{Aubert:2005ye}
\bibinfo{author}{\bibfnamefont{B.}~\bibnamefont{Aubert}} \bibnamefont{et~al.}
  (\bibinfo{collaboration}{\babar} Collaboration), \bibinfo{journal}{Phys. Rev.
  Lett.} \textbf{\bibinfo{volume}{95}}, \bibinfo{pages}{041802}
  (\bibinfo{year}{2005}).

\bibitem[{\citenamefont{Aubert et~al.}(2006)}]{Aubert:2005wa}
\bibinfo{author}{\bibfnamefont{B.}~\bibnamefont{Aubert}} \bibnamefont{et~al.}
  (\bibinfo{collaboration}{\babar} Collaboration), \bibinfo{journal}{Phys. Rev.
  Lett.} \textbf{\bibinfo{volume}{96}}, \bibinfo{pages}{041801}
  (\bibinfo{year}{2006}).

\bibitem[{\citenamefont{Hayasaka et~al.}(2008)}]{Hayasaka:2007vc}
\bibinfo{author}{\bibfnamefont{K.}~\bibnamefont{Hayasaka}} \bibnamefont{et~al.}
  (\bibinfo{collaboration}{Belle} Collaboration), \bibinfo{journal}{Phys. Lett.
  B} \textbf{\bibinfo{volume}{666}}, \bibinfo{pages}{16}
  (\bibinfo{year}{2008}).

\bibitem[{\citenamefont{Raidal et~al.}(2008)}]{Raidal:2008jk}
\bibinfo{author}{\bibfnamefont{M.}~\bibnamefont{Raidal}} \bibnamefont{et~al.}
  (\bibinfo{year}{2008}), \eprint{hep-ph/0801.1826}.

\bibitem[{ftn()}]{ftnt1}
\eprint{Charge conjugate decays are implicitly included.}

\bibitem[{\citenamefont{Ilakovac}(2000)}]{Ilakovac:1999md}
\bibinfo{author}{\bibfnamefont{A.}~\bibnamefont{Ilakovac}},
  \bibinfo{journal}{Phys. Rev. D} \textbf{\bibinfo{volume}{62}},
  \bibinfo{pages}{036010} (\bibinfo{year}{2000}).

\bibitem[{\citenamefont{Saha and Kundu}(2002)}]{Saha:2002kt}
\bibinfo{author}{\bibfnamefont{J.~P.} \bibnamefont{Saha}} \bibnamefont{and}
  \bibinfo{author}{\bibfnamefont{A.}~\bibnamefont{Kundu}},
  \bibinfo{journal}{Phys. Rev. D} \textbf{\bibinfo{volume}{66}},
  \bibinfo{pages}{054021} (\bibinfo{year}{2002}).

\bibitem[{\citenamefont{Miyazaki et~al.}(2006)}]{Miyazaki:2006sx}
\bibinfo{author}{\bibfnamefont{Y.}~\bibnamefont{Miyazaki}} \bibnamefont{et~al.}
  (\bibinfo{collaboration}{Belle} Collaboration), \bibinfo{journal}{Phys. Lett.
  B} \textbf{\bibinfo{volume}{639}}, \bibinfo{pages}{159}
  (\bibinfo{year}{2006}).

\bibitem[{\citenamefont{Aubert et~al.}(2002)}]{Aubert:2001tu}
\bibinfo{author}{\bibfnamefont{B.}~\bibnamefont{Aubert}} \bibnamefont{et~al.}
  (\bibinfo{collaboration}{\babar} Collaboration), \bibinfo{journal}{Nucl.
  Instrum. Meth. A} \textbf{\bibinfo{volume}{479}}, \bibinfo{pages}{1}
  (\bibinfo{year}{2002}).

\bibitem[{\citenamefont{Menges}(2006)}]{Menges:2006xk}
\bibinfo{author}{\bibfnamefont{W.}~\bibnamefont{Menges}},
  \bibinfo{journal}{IEEE Nucl. Sci. Symp. Conf. Rec.}
  \textbf{\bibinfo{volume}{5}}, \bibinfo{pages}{1470} (\bibinfo{year}{2006}).

\bibitem[{\citenamefont{Banerjee et~al.}(2008)\citenamefont{Banerjee, Pietrzyk,
  Roney, and Was}}]{Banerjee:2007is}
\bibinfo{author}{\bibfnamefont{S.}~\bibnamefont{Banerjee}},
  \bibinfo{author}{\bibfnamefont{B.}~\bibnamefont{Pietrzyk}},
  \bibinfo{author}{\bibfnamefont{J.~M.} \bibnamefont{Roney}} \bibnamefont{and}
  \bibinfo{author}{\bibfnamefont{Z.}~\bibnamefont{Was}},
  \bibinfo{journal}{Phys. Rev. D} \textbf{\bibinfo{volume}{77}},
  \bibinfo{pages}{054012} (\bibinfo{year}{2008}).

\bibitem[{\citenamefont{Jadach et~al.}(2000)\citenamefont{Jadach, Ward, and
  Was}}]{Jadach:1999vf}
\bibinfo{author}{\bibfnamefont{S.}~\bibnamefont{Jadach}},
  \bibinfo{author}{\bibfnamefont{B.~F.~L.} \bibnamefont{Ward}}
  \bibnamefont{and} \bibinfo{author}{\bibfnamefont{Z.}~\bibnamefont{Was}},
  \bibinfo{journal}{Comput. Phys. Commun.} \textbf{\bibinfo{volume}{130}},
  \bibinfo{pages}{260} (\bibinfo{year}{2000}).

\bibitem[{\citenamefont{Ward et~al.}(2003)\citenamefont{Ward, Jadach, and
  Was}}]{Ward:2002qq}
\bibinfo{author}{\bibfnamefont{B.~F.~L.} \bibnamefont{Ward}},
  \bibinfo{author}{\bibfnamefont{S.}~\bibnamefont{Jadach}} \bibnamefont{and}
  \bibinfo{author}{\bibfnamefont{Z.}~\bibnamefont{Was}},
  \bibinfo{journal}{Nucl. Phys. Proc. Suppl.} \textbf{\bibinfo{volume}{116}},
  \bibinfo{pages}{73} (\bibinfo{year}{2003}).

\bibitem[{\citenamefont{Jadach et~al.}(1993)\citenamefont{Jadach, Was, Decker,
  and Kuhn}}]{Jadach:1993hs}
\bibinfo{author}{\bibfnamefont{S.}~\bibnamefont{Jadach}},
  \bibinfo{author}{\bibfnamefont{Z.}~\bibnamefont{Was}},
  \bibinfo{author}{\bibfnamefont{R.}~\bibnamefont{Decker}} \bibnamefont{and}
  \bibinfo{author}{\bibfnamefont{J.~H.} \bibnamefont{Kuhn}},
  \bibinfo{journal}{Comput. Phys. Commun.} \textbf{\bibinfo{volume}{76}},
  \bibinfo{pages}{361} (\bibinfo{year}{1993}).

\bibitem[{\citenamefont{Barberio and Was}(1994)}]{Barberio:1993qi}
\bibinfo{author}{\bibfnamefont{E.}~\bibnamefont{Barberio}} \bibnamefont{and}
  \bibinfo{author}{\bibfnamefont{Z.}~\bibnamefont{Was}},
  \bibinfo{journal}{Comput. Phys. Commun.} \textbf{\bibinfo{volume}{79}},
  \bibinfo{pages}{291} (\bibinfo{year}{1994}).

\bibitem[{\citenamefont{Lange}(2001)}]{Lange:2001uf}
\bibinfo{author}{\bibfnamefont{D.~J.} \bibnamefont{Lange}},
  \bibinfo{journal}{Nucl. Instrum. Meth. A} \textbf{\bibinfo{volume}{462}},
  \bibinfo{pages}{152} (\bibinfo{year}{2001}).

\bibitem[{\citenamefont{Sjostrand et~al.}(2006)\citenamefont{Sjostrand, Mrenna,
  and Skands}}]{Sjostrand:2006za}
\bibinfo{author}{\bibfnamefont{T.}~\bibnamefont{Sjostrand}},
  \bibinfo{author}{\bibfnamefont{S.}~\bibnamefont{Mrenna}} \bibnamefont{and}
  \bibinfo{author}{\bibfnamefont{P.}~\bibnamefont{Skands}},
  \bibinfo{journal}{JHEP} \textbf{\bibinfo{volume}{05}}, \bibinfo{pages}{026}
  (\bibinfo{year}{2006}).

\bibitem[{\citenamefont{Agostinelli et~al.}(2003)}]{Agostinelli:2002hh}
\bibinfo{author}{\bibfnamefont{S.}~\bibnamefont{Agostinelli}}
  \bibnamefont{et~al.} (\bibinfo{collaboration}{{GEANT}} Collaboration),
  \bibinfo{journal}{Nucl. Instrum. Meth. A} \textbf{\bibinfo{volume}{506}},
  \bibinfo{pages}{250} (\bibinfo{year}{2003}).

\bibitem[{\citenamefont{Golonka et~al.}(2006)}]{Golonka:2003xt}
\bibinfo{author}{\bibfnamefont{P.}~\bibnamefont{Golonka}} \bibnamefont{et~al.},
  \bibinfo{journal}{Comput. Phys. Commun.} \textbf{\bibinfo{volume}{174}},
  \bibinfo{pages}{818} (\bibinfo{year}{2006}).

\bibitem[{\citenamefont{Yao et~al.}(2006)}]{Yao:2006px}
\bibinfo{author}{\bibfnamefont{W.~M.} \bibnamefont{Yao}} \bibnamefont{et~al.}
  (\bibinfo{collaboration}{Particle Data Group} Collaboration),
  \bibinfo{journal}{J. Phys. G} \textbf{\bibinfo{volume}{33}},
  \bibinfo{pages}{1} (\bibinfo{year}{2006}).

\bibitem[{\citenamefont{Brandt et~al.}(1964)\citenamefont{Brandt, Peyrou,
  Sosnowski, and Wroblewski}}]{Brandt:1964sa}
\bibinfo{author}{\bibfnamefont{S.}~\bibnamefont{Brandt}},
  \bibinfo{author}{\bibfnamefont{C.}~\bibnamefont{Peyrou}},
  \bibinfo{author}{\bibfnamefont{R.}~\bibnamefont{Sosnowski}} \bibnamefont{and}
  \bibinfo{author}{\bibfnamefont{A.}~\bibnamefont{Wroblewski}},
  \bibinfo{journal}{Phys. Lett.} \textbf{\bibinfo{volume}{12}},
  \bibinfo{pages}{57} (\bibinfo{year}{1964}).

\bibitem[{\citenamefont{Junk}(1999)}]{Junk:1999kv}
\bibinfo{author}{\bibfnamefont{T.}~\bibnamefont{Junk}}, \bibinfo{journal}{Nucl.
  Instrum. Meth. A} \textbf{\bibinfo{volume}{434}}, \bibinfo{pages}{435}
  (\bibinfo{year}{1999}).

\bibitem[{Jam(2000)}]{James:2000et}
\emph{\bibinfo{title}{Workshop on confidence limits, CERN, Geneva, Switzerland,
  17-18 Jan 2000: Proceedings}} (\bibinfo{year}{2000}),
  \bibinfo{note}{{CERN-2000-005}}.

\bibitem[{\citenamefont{Conrad et~al.}(2003)\citenamefont{Conrad, Botner,
  Hallgren, and Perez de~los Heros}}]{Conrad:2002kn}
\bibinfo{author}{\bibfnamefont{J.}~\bibnamefont{Conrad}},
  \bibinfo{author}{\bibfnamefont{O.}~\bibnamefont{Botner}},
  \bibinfo{author}{\bibfnamefont{A.}~\bibnamefont{Hallgren}} \bibnamefont{and}
  \bibinfo{author}{\bibfnamefont{C.}~\bibnamefont{Perez de~los Heros}},
  \bibinfo{journal}{Phys. Rev. D} \textbf{\bibinfo{volume}{67}},
  \bibinfo{pages}{012002} (\bibinfo{year}{2003}).

\end{thebibliography}
\end{document}